\documentclass[preprint,showpacs,preprintnumbers,amsmath,amssymb,showkeys]{revtex4}

\usepackage{graphicx}
\usepackage{dcolumn}
\usepackage{bm}

\begin{document}

\preprint{}

\title{A polarized beam splitter using an anisotropic medium slab}

\author{Hailu Luo}\thanks{Author to whom correspondence should be addressed.
E-mail: hailuluo@sohu.com}
\author{Weixing Shu}
\author{Fei Li}
\author{Zhongzhou Ren}

\affiliation{Department of Physics, Nanjing University, Nanjing
210008, China}
\date{\today}

\begin{abstract}
The propagation of electromagnetic waves in the anisotropic medium
with a single-sheeted hyperboloid dispersion relation is
investigated. It is found that in such an anisotropic medium E-
and H-polarized waves have the same dispersion relation, while E-
and H-polarized waves exhibit opposite amphoteric refraction
characteristics. E- (or H-) polarized waves are positively
refracted whereas H- (or E-) polarized waves are negatively
refracted at the interface associated with the anisotropic medium.
By suitably using the properties of anomalous refraction in the
anisotropic medium it is possible to realize a very simple and
very efficient beam splitter to route the light. It is shown that
the splitting angle and the splitting distance between E- and H-
polarized beam is the function of anisotropic parameters, incident
angle and slab thickness.
\end{abstract}

\maketitle

\section{Introduction}\label{Introduction}
Polarization beam splitter is an important device in optical
systems, such as polarization-independent optical isolators and
optical switches~\cite{Born1999,Yariv1984}. A conventional
polarization beam splitter is made of a nonmagnetic anisotropic
crystal or a multi-layer transparent
material~\cite{Shiraishi1991,Muro1998,Shiraishi1998}. The
separation between E- and H- polarized beams produced by these
conventional methods is typically limited by the small splitting
angle. While a large beam splitting angle and splitting distance
are preferable for practical applications, especially in the field
of optical communication systems.

In general, E- and H- polarized waves propagate in different
directions in an anisotropic medium. For a regular anisotropic
medium, all tensor elements of permittivity
$\boldsymbol{\varepsilon}$ and permeability $\boldsymbol{\mu}$
should be positive. The recent advent of a new class of material
with negative permittivity and permeability has attained
considerable
attention~\cite{Veselago1968,Smith2000,Shelby2001,Parazzoli2003,Houck2003}.
Recently, Lindell {\it et al.} \cite{Lindell2001} have shown that
anomalous negative refraction can occur at an interface associated
with an anisotropic media, which does not necessarily require that
all tensor elements of $\boldsymbol{\varepsilon}$ and
$\boldsymbol{\mu}$ have negative values. The studies of such
anisotropic media have recently received much interest and
attention~\cite{Hu2002,Zhou2003,Smith2003,Thomas2005,Luo2005,Shen2005,Luo2006a,Depine2006a,Depine2006b}.
Although E- and H-polarized waves can present the amphoteric
refraction in conventional nonmagnetic anisotropic
crystal~\cite{Luo2005}, the splitting angle is very small. A
question naturally arise: whether there exists a large splitting
angle and splitting distance, when E- and H-polarized waves
propagating in certain anisotropic media.

In this paper we investigate the propagation of electromagnetic
waves in an anisotropic material with a single-sheeted hyperboloid
dispersion relation. We show that E- (or H-) polarized beam is
positively refracted whereas H- (or E-) polarized beam is
negatively refracted at the interface associated with the
anisotropic media. There exists both a large splitting angle and
splitting distance, when E- and H-polarized waves in the special
anisotropic media. We present a design of polarization beam
splitters based on the anomalous refraction in the anisotropic
medium. To match the boundary conditions, the material parameters
of the anisotropic medium can be designed.

\section{Dispersion relations of an anisotropic media}\label{sec2}
Before we consider the beam splitter structure, we first analyze
the dispersion relation of the anisotropic medium. For anisotropic
materials one or both of the permittivity and permeability are
second-rank tensors~\cite{Chen1983,Kong1990}. To simplify the
proceeding analysis, we assume the permittivity and permeability
tensors are simultaneously diagonalizable:
\begin{eqnarray}
\boldsymbol{\varepsilon}=\left(
\begin{array}{ccc}
\varepsilon_x  &0 &0 \\
0 & \varepsilon_y &0\\
0 &0 & \varepsilon_z
\end{array}
\right), ~~\boldsymbol{\mu}=\left(
\begin{array}{ccc}
\mu_x  &0 &0 \\
0 & \mu_y &0\\
0 &0 & \mu_z
\end{array}
\right),\label{matrix}
\end{eqnarray}
where $\varepsilon_i$ and $\mu_i$  are the permittivity and
permeability constants in the principal coordinate system
($i=x,y,z$).

We choose the $z$ axis to be normal to the interface, and the $x$,
$y$ axes locate at the plane of the interface. We assume plane
wave with frequency $\omega$ incident from isotropic media into
anisotropic media. In isotropic media the accompanying dispersion
relation has the familiar form
\begin{equation}
 k_{x}^2+ k_{y}^2+k_{z}^2= \varepsilon_I \mu_I\frac{\omega^2}{c^2}. \label{D1}
\end{equation}
Here $ k_i$ is the $i$ component of the propagating wave vector
and $c$ is the speed of light in vacuum. $\varepsilon_I$ and
$\mu_I$ are the permittivity and permeability, respectively.

\begin{figure}
\includegraphics[width=8cm]{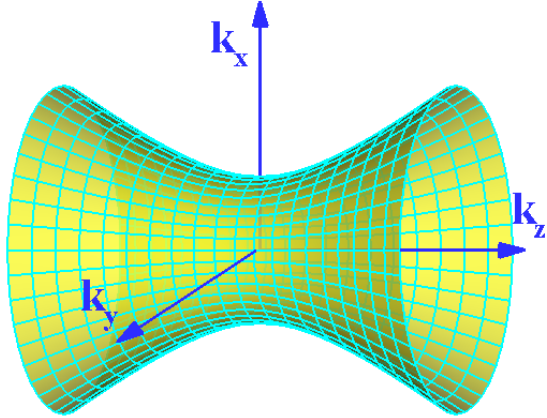}
\caption{\label{Fig1} E- and H-polarized waves have the same
single-sheeted hyperboloid dispersion relation.}
\end{figure}

In the uniaxially anisotropic media, the E- and H- polarized
incident waves have the dispersion
relation~\cite{Chen1983,Shen2005}
\begin{equation}
\frac{ q_{x}^2}{\varepsilon_y \mu_z}+\frac{q_{y}^2}{\varepsilon_x
\mu_z}+\frac{q_{z}^2}{\varepsilon_y
\mu_x}-\frac{\omega^2}{c^2}=0,\label{D1}
\end{equation}
\begin{equation}
\frac{ q_{x}^2}{\varepsilon_z \mu_y}+\frac{q_{y}^2}{\varepsilon_z
\mu_x}+\frac{q_{z}^2}{\varepsilon_x
\mu_y}-\frac{\omega^2}{c^2}=0.\label{D2}
\end{equation}
Here $q_{i}$ represents the $i$ component of transmitted
wave-vector. If the permittivity and permeability constants
satisfy the relation:
\begin{equation}
\frac{\varepsilon_x }{\mu_x}=\frac{\varepsilon_y
}{\mu_y}=\frac{\varepsilon_z }{\mu_z}=C~~~(C<0), \label{D1}
\end{equation}
where $C$ is a constant, then the E- and H- polarized waves have
the same dispersion relation~\cite{Shen2005,Luo2006b}
\begin{equation}
\frac{ q_{x}^2}{\varepsilon_z \mu_y}+\frac{q_{y}^2}{\varepsilon_z
\mu_x}+\frac{q_{z}^2}{\varepsilon_y \mu_x}= \frac{\omega^2}{c^2},
\label{D2}
\end{equation}

Based on the dispersion relation one can find that the wave-vector
surface is a single-sheeted hyperboloid. It should be mentioned
that the propagation character in quasiisotropic media $C>0$ has
been discussed in our previous work. We have shown that E- and H-
polarized waves have the same propagation character in
quasiisotropic medium~\cite{Luo2006a}. In present work, we want to
enquire whether E- and H- polarized waves have the same
propagation feature.

\section{Positive and negative refraction}\label{sec3}
In this section, we shall answer the question asked in the above
section. The $z$-component of the wave vector can be found by the
solution of Eq. (\ref{D2}), which yields
\begin{equation}
 q_z = \sigma\sqrt {\varepsilon_y \mu_x\frac{\omega^2}{c^2}-
\left(\frac{\varepsilon_y \mu_x }{\varepsilon_z
\mu_y}q_{x}^2+\frac{\varepsilon_y }{\varepsilon_z
}q_{y}^2\right)}, \label{qz}
\end{equation}
where $\sigma=+1$ or $\sigma=-1$. The choice of the sign ensures
that light power propagates away from the surface to the $+z$
direction.

Without loss of generality, we assume the wave vector locate in
the $x-z$  plane ($k_y=q_y=0$). The incident angle of light is
given by
\begin{equation}
\theta_I =\tan^{-1}\left[\frac{k_x}{k_{z}}\right],
\end{equation}
In principle the occurrence of refraction requires that the $z$
component of the wave vector of the refracted waves must be real.
Then the incident angle must satisfy the following inequality:
\begin{equation}
\frac{\varepsilon_y \mu_x }{\varepsilon_z
\mu_y}q_{x}^2>\varepsilon_y \mu_x\frac{\omega^2}{c^2}.\label{IC}
\end{equation}
Based on the boundary condition, the tangential components of the
wave vectors must be continuous
\begin{equation}
q_x=k_x=\frac{\sqrt{\varepsilon_I \mu_I}\omega}{c}\sin
\theta_I.\label{BC}
\end{equation}
Substituting Eq.~(\ref{BC}) into Eq.~(\ref{IC}), one can obtain E-
and H- polarized waves have the same critical angle
\begin{equation}
\theta_C^{E}=\theta_C^{H}=\sin^{-1}\left[\sqrt{\frac{\varepsilon_z
\mu_y}{\varepsilon_I \mu_I}}\right].\label{CA}
\end{equation}

\begin{figure}
\includegraphics[width=8cm]{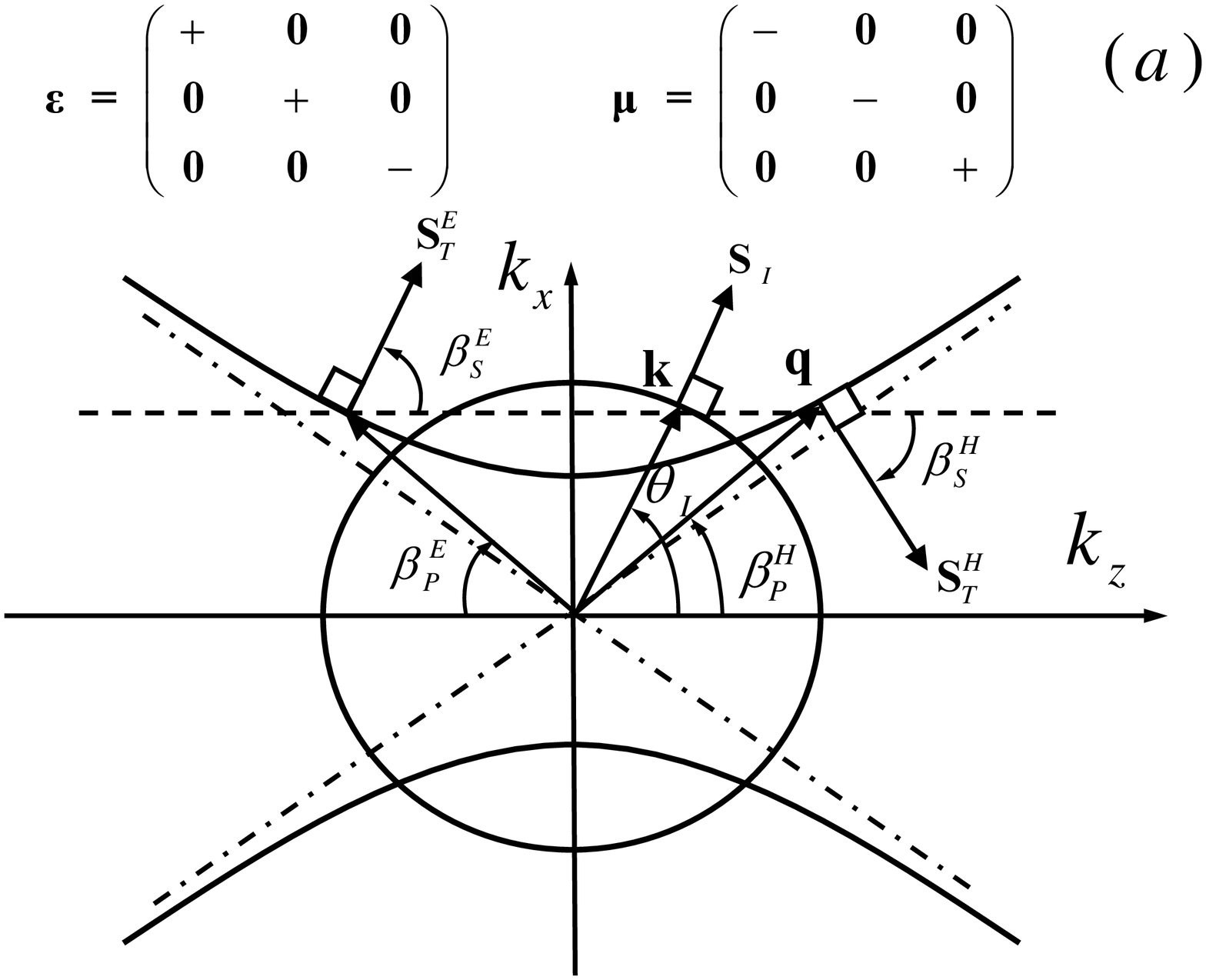}
\includegraphics[width=8cm]{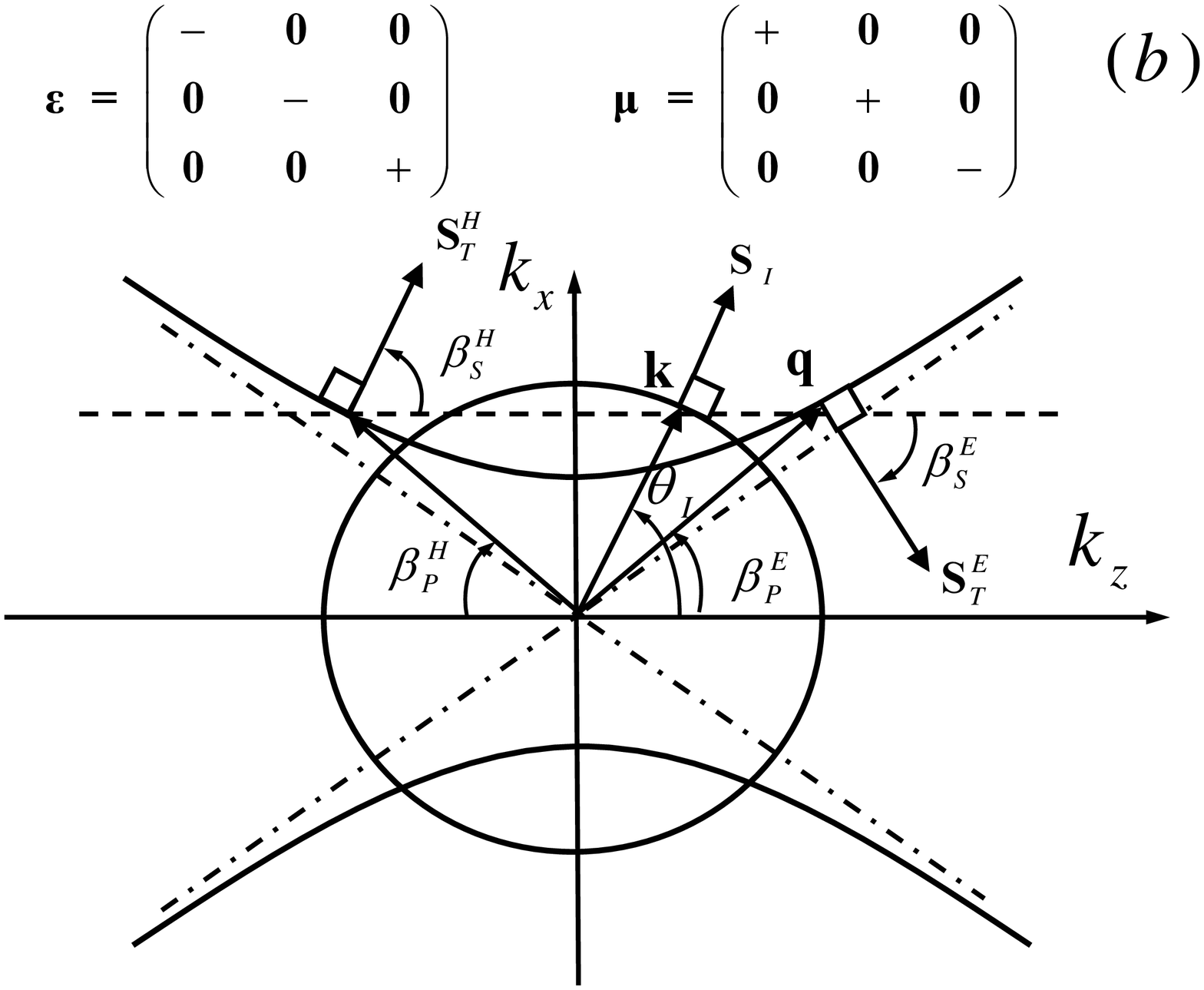}
\caption{\label{Fig2} The circle and single-sheeted hyperbola
represent the dispersion relations of the isotropic medium and the
anisotropic medium, respectively. The incident wave vector ${\bf
k}$ is parallel to the Poynting vector ${\bf S}_I$ in the
isotropic medium. Because the anisotropy, ${\bf S}_T$ must lie
normal to the frequency contour. The single-sheeted hyperboloid
wave-vector surface has two types: (a)
$\beta_P^{E}=\beta_P^{H}>0$, $\beta_S^{E}=\beta_S^{H}<0$, (b)
$\beta_P^{E}=\beta_P^{H}<0$, $\beta_S^{E}=\beta_S^{H}>0$.}
\end{figure}

The refractive angle of the transmitted wave vector or phase of E-
and H-polarized waves can be written as
\begin{equation}
\beta_P^{E}=\tan^{-1}\left[\frac{q_x^{E}}{q_{z}^{E}}\right],~~~
\beta_P^{H}=\tan^{-1}\left[\frac{q_x^{H}}{q_{z}^{H}}\right].\label{AP}
\end{equation}

It should be noted that the actual direction of light is defined
by the time-averaged Poynting vector ${\bf S} =\frac{1}{2} {\bf
Re}({\bf E}^\ast\times \bf{H})$. For E- polarized incident waves,
the Poynting vector is given by
\begin{equation}
{\bf S}_T^{E}=Re \left[\frac{E_0^2 q_x^{E}}{2 \omega \mu_z}{\bf
e}_x+\frac{E_0^2 q_z^{E}}{2\omega\mu_x}{\bf e}_z\right].\label{SE}
\end{equation}
For H-polarized incident waves, the transmitted Poynting vector is
given by
\begin{equation}
{\bf S}_T^{H}=Re\left[\frac{H_0^2 q_x^{H}}{ 2
\omega\varepsilon_z}{\bf e}_x+\frac{H_0^2
q_z^{H}}{2\omega\varepsilon_x}{\bf e}_z\right].\label{SH}
\end{equation}
The refractive angle of Poynting vector of E- and H- polarized
incident waves can be obtained as
\begin{equation}
\beta_S^{E}= \tan^{-1}\left[\frac{S_{Tx}^{E}}{S_{Tz}^{E}}\right],
~~~ \beta_S^{H}=
\tan^{-1}\left[\frac{S_{Tx}^{H}}{S_{Tz}^{H}}\right].\label{AS}
\end{equation}

For the purpose of illustration, the frequency contour will be
used to determine the refracted waves as shown in Fig.~\ref{Fig2}.
From the boundary condition $q_x=k_x$, we can obtain two
possibilities for the refracted wave vector. Energy conservation
requires that the $z$ component of poynting vector must propagates
away from the interface, for instance, $q_z^{E}/\mu_x>0$ and
$q_z^{H}/\varepsilon_x>0$. Then the sign of $q_z^E$ and $q_z^H$
can be determined easily. The corresponding Poynting vector should
be drawn perpendicularly to the dispersion contour. Thus we can
obtain two possibilities (inward or outward), while only the
Poynting vector with $S_{Tz}>0$ is causal.

As can be seen from Fig.~\ref{Fig2}, the wave-vector surface is a
single-sheeted hyperbola. This medium
has two types:\\
(I)~  For the case of $\varepsilon_x>0$, $\varepsilon_y>0$ and
$\varepsilon_z<0$, the refraction diagram is plotted in
Fig.~\ref{Fig2}a. For E-polarized incident waves ${\bf
k}_z\cdot{\bf q}_{z}^{E}<0$, ${\bf k}_x\cdot{\bf S}_{T}^{E}>0$.
For H-polarized incident waves ${\bf k}_z\cdot{\bf q}_{z}^{H}>0$,
${\bf k}_x\cdot{\bf S}_{T}^{H}<0$. From Eqs.~(\ref{SE}) and
(\ref{SH}), if $\varepsilon_x/\mu_x=\varepsilon_z/\mu_z=C<0$,
$S^{E}_{Tz}$  and $S^{H}_{Tz}$ have the same sign, while
$S^{E}_{Tx}$  and $S^{H}_{Tx}$ are always in the opposite sign,
i.e.
\begin{equation}
\beta_P^{E}=-\beta_P^{H}<0,~~~\beta_S^{E}=-\beta_S^{H}>0.\label{II1}
\end{equation}
(II)  For the case of $\varepsilon_x<0$, $\varepsilon_y<0$ and
$\varepsilon_z>0$, the refraction diagram is plotted in
Fig.~\ref{Fig2}b. For E-polarized incident waves ${\bf
k}_z\cdot{\bf q}_{z}^{E}>0$, ${\bf k}_x\cdot{\bf S}_{T}^{E}<0$.
For H-polarized incident waves ${\bf k}_z\cdot{\bf q}_{z}^{H}<0$,
${\bf k}_x\cdot{\bf S}_{T}^{H}>0$.  From Eqs. (\ref{SE}) and
(\ref{SH}) we can get
\begin{equation}
\beta_P^{E}=-\beta_P^{H}>0,~~~\beta_S^{E}=-\beta_S^{H}<0.\label{II2}
\end{equation}
We thus conclude that E- and H-polarized waves have the same
dispersion relation while E- (or H-) polarized beam is positively
refracted whereas H- (or E-) polarized beam is negatively
refracted in such anisotropic media.

\begin{figure}
\includegraphics[width=10cm]{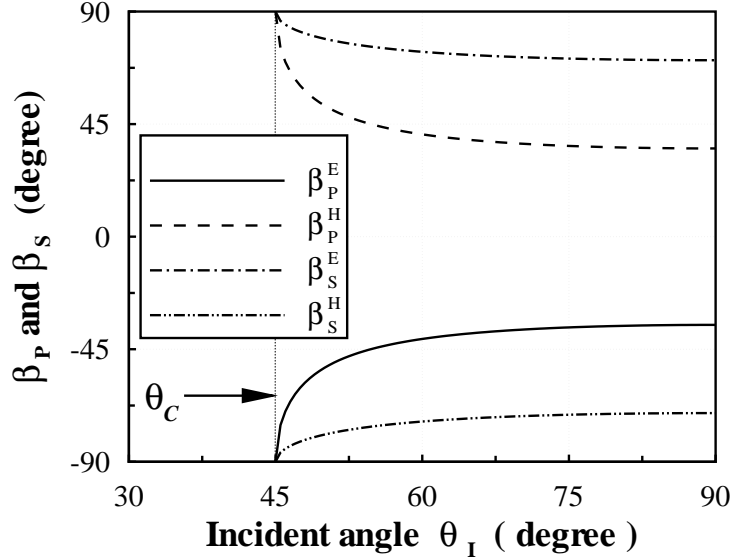}
\caption{\label{Fig3} The variations of the refractive angle
$\beta_S$ for energy flow and the transmitted wave vector angle
$\beta_P$ as a function of the incident angle $\theta_I$,  The
E-polarized beam undergoes a positive refraction, while the energy
flow of H-polarized beam undergoes negative refraction.}
\end{figure}

In following analysis we are interested in type I, in which the
E-polarized waves are positively refracted whereas the H-polarized
waves are negatively refracted. The refractive angles of Poynting
vector and wave vector are plotted in Fig.~\ref{Fig3}.  For the
purpose of illustration, We choose some simple anisotropic
parameters, i.e. $\varepsilon=2$, $\varepsilon_y=1$,
$\varepsilon_z=0.5$ and $C=-1$. It should be mentioned that the
parameters can be effectively modelled in periodic wires and rings
structure \cite{Smith2003,Thomas2005}.

The above analysis suggest that a large splitting angle  can be
obtained by tuning the anisotropic parameters. The splitting angle
between E- and H-polarized waves can be defined as
\begin{equation}
\Phi=\beta_S^{E}-\beta_S^{H}.\label{AS}
\end{equation}
If the waves incident at $\theta=60^\circ$, the splitting angle
will reach a large value $\Phi=148^\circ$ as shown in
Fig~\ref{Fig3}. A question naturally arise: why do we not use the
special medium to construct an efficient splitter?

\section{Waves incident at Brewster angle }\label{sec3}
\begin{figure}
\includegraphics[width=10cm]{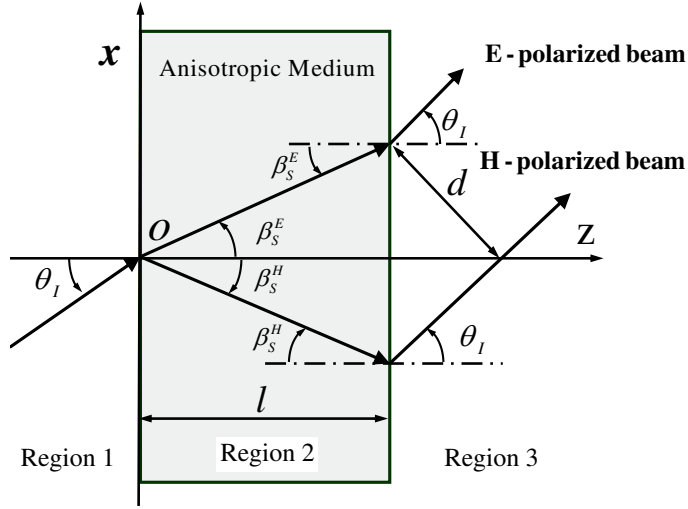}
\caption{\label{Fig4} Schematic diagram illustrating the polarized
beams splitter}
\end{figure}

From the analysis of the previous section we know that E-polarized
beam is positively refracted whereas H-polarized beam is
negatively refracted by the anisotropic slab. The interesting
properties allow us to introduce the potential device acting as a
polarizing beam splitter. The optical beam splitter consists of an
anisotropic medium slab as shown in Fig.~\ref{Fig4}.

The media parameters of anisotropic slab can be tuned to meet the
requirements, for example, the reflection coefficient of a
E-polarized wave equals to that of a H-polarized wave. Based on
the boundary conditions, the reflection and the transmission
coefficients can be obtained~\cite{Luo2006b}. For E-polarized
incident waves, one can obtain the following expression for the
reflection and transmission coefficients
\begin{equation}
R_E=\frac{\mu_x k_z-\mu_I q_z^{E}}{\mu_x  k_z+ \mu_I
q_z^{E}},~~~T_E = \frac{2 \mu_x k_z}{\mu_x  k_z+\mu_I
q_z^{E}}.\label{RETE}
\end{equation}
For H-polarized incident waves, the reflection and transmission
coefficients can be obtained similarly as
\begin{equation}
R_H=\frac{\varepsilon_x k_z-\varepsilon_I q_z^{H}}{\varepsilon_x
k_z+\varepsilon_I q_z^{H}},~~~T_H = \frac{2 \varepsilon_x
k_z}{\varepsilon_x k_z+\varepsilon_I q_z^{H}}.\label{RHTH}
\end{equation}

Mathematically the Brewster angles can be obtained from $T_E=0$
and $T_H=0$. For E-polarized incident waves if the anisotropic
parameters satisfy the relation
\begin{equation}
0<\frac{ \mu_z(\varepsilon_y \mu_I - \varepsilon_I
\mu_x)}{\varepsilon_I (\mu_I^2 -\mu_x \mu_z ) }< 1,\label{EBC}
\end{equation}
the Brewster angle can be expressed as
\begin{equation}
\theta_B^{E}=\sin^{-1}\left[\sqrt{\frac{ \mu_z(\varepsilon_y \mu_I
- \varepsilon_I \mu_x)}{\varepsilon_I (\mu_I^2 -\mu_x \mu_z ) }}
\right].\label{EB}
\end{equation}
For H-polarized waves if the anisotropic parameters satisfy by the
relation
\begin{equation}
0<\frac{ \varepsilon_z( \varepsilon_I \mu_y - \varepsilon_x\mu_I
)}{\mu_I (\varepsilon_I^2- \varepsilon_x \varepsilon_z)}<
1,\label{HBC}
\end{equation}
the Brewster angle can be written in the form
\begin{equation}
\theta_B^{H}=\sin^{-1}\left[\sqrt{\frac{ \varepsilon_z(
\varepsilon_I \mu_y - \varepsilon_x\mu_I  )}{\mu_I
(\varepsilon_I^2- \varepsilon_x \varepsilon_z) }}
\right].\label{HB}
\end{equation}

It should be mentioned that E- and H-polarized waves may exhibit a
Brewster angle simultaneously, which depends on the choice of the
anisotropic parameters. Clearly, if one seeks a solution
satisfying Eq.~(\ref{EB}) and Eq.~(\ref{HB}), the only possibility
is
\begin{equation}
\frac{\varepsilon_x }{\mu_x}=\frac{\varepsilon_y
}{\mu_y}=\frac{\varepsilon_z }{\mu_z}=-\frac{\varepsilon_I
}{\mu_I}, \label{D1}
\end{equation}
one can obtain an interesting features: E- and H-polarized waves
will exhibit the same reflection and transmission, namely,
$R_E=R_H $ and $T_E=T_H$.

\begin{figure}
\includegraphics[width=10cm]{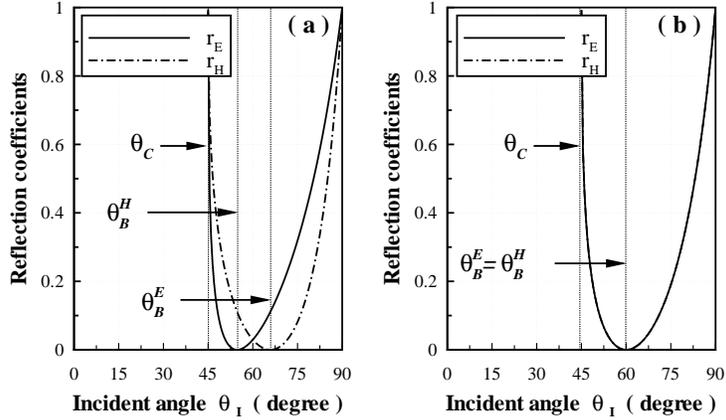}
\caption{\label{Fig5} The reflection coefficients of E- and
H-polarized waves as functions of the incident angle $\theta_I$.
E- and H-polarized waves exhibit a Brewster angle simultaneously:
(a) $\theta_B^{E}\neq\theta_B^{H}$, (b)
$\theta_B^{E}=\theta_B^{H}$.}
\end{figure}

The reflection coefficients of E- and H-polarized waves are
plotted in Fig.~\ref{Fig5}. We choose some simple parameters for
the purpose of illustration, i.e., $C=-1$. In Fig.~\ref{Fig5}a E-
and H-polarized incident waves exhibit a Brewster angle,
simultaneously. When the incident angle equal to the Brewster
angle, the incident waves will exhibit oblique total transmission.
In Fig.~\ref{Fig5}b E- and H-polarized waves exhibit the same
Brewster angle. We can find E- and H-polarized waves will exhibit
oblique total transmission at same indigent angle.

To construct an efficient splitter, we wish that E- and
H-polarized waves can totally transmit thought the anisotropic
slab. We thus choose the incident angle equal to the Brewster
angle $\theta_I$=$\theta_B^{E}$=$\theta_B^{H}$, the reflects of E-
and H- polarized waves are completely absent.

\section{Polarizing Beam Splitter based on the anomalous refraction}
The splitting distance (or walk-off distance) between E- and H-
polarized beam is the functions of anisotropic parameters, the
incident angle and the slab thickness. The splitting distance can
be easily obtained as
\begin{equation}
d=2\cos \theta_I \tan \beta_S^{E} l,\label{ASEH}
\end{equation}
where $d$ is the splitting distance and  $l$ is the thickness of
slab.

To obtain a better physical picture of beam splitter, a modulated
Gaussian beam of finite width can be constructed. Following the
method outlined by Lu {\it et al.}~\cite{Lu2004}, let us consider
a modulated beam incident from free space
\begin{equation}
E_1(x, z) = \int_{-\infty}^{+\infty}d k_\perp f( k_\perp)
\exp[i({\bf k}_0+{\bf k}_\perp) \cdot {\bf r}- i \omega_0  t],
\label{FI}
\end{equation}
where ${\bf k}_\perp$ is perpendicular to ${\bf k}_0$ and
$\omega_0=c k_0$. A general incident wave vector is written as
${\bf k}={\bf k}_0+{\bf k}_\perp $. we assume its Gaussian weight
is
\begin{equation}
f (k_\perp) = \frac{w_0}{\sqrt{\pi}} \exp [- w_0^2 k_\perp ],
\label{fk}
\end{equation}
where $w_0$ is the spatial extent of the incident beam. We want
the modulated Gaussian beam to be aligned with the incident
direction defined by the vector ${\bf k}_0=k_0 \cos \theta_I {\bf
e}_x+ k_0 \sin \theta_I {\bf e}_z$, which makes the incident angle
equal to the Brewster angle.

\begin{figure}
\includegraphics[width=12cm]{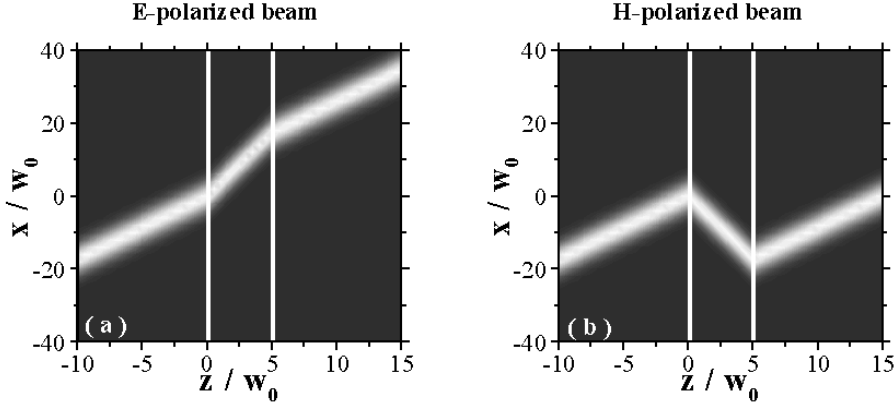}
\caption{\label{Fig6} Observed polarization-splitting
characteristics and their intensity distributions. (a) E-polarized
beam is positively refracted, (b) H-polarized beam is negatively
refracted. The anisotropic parameters are same as those used in
Fig.~\ref{Fig2}. The splitting distance between the two beam peaks
is $17.3 w_0$.}
\end{figure}

Matching the boundary conditions for each $k$ component at $z=0$
gives the complex field in the form
\begin{equation}
E_{2}(x, z) = \int_{-\infty}^{+\infty}d k_\perp f( k_\perp )T_1
({\bf k}) \exp[i({\bf q}\cdot {\bf r}-\omega_0 t)].\label{FII}
\end{equation}
where $T_1({\bf k})$ is the transmission coefficient of the
modulated Gaussian beam at the first interface. The transmission
coefficient can obtain a good approximation to simply evaluate
this quantity at ${\bf k}_0$. We set the waves are incident at the
Brewster angle, then the transmission coefficient of the modulated
beam is simply given by $T_1({\bf k}_0)=1$. The normal component
of refracted wave vector $q_z$ can be expanded in a Taylor series
to first order in ${\bf k}_0 $ to obtain a better approximation
\begin{equation}
q_z^E({\bf k}) =q_z^E({\bf k}_0 )+({\bf k}-{\bf k}_0 )\cdot
\frac{\partial q_z^E({\bf k})}{\partial {\bf k}} \bigg|_{{\bf
k}_0}.\label{te}
\end{equation}
\begin{equation}
q_z^H({\bf k}) =q_z^H({\bf k}_0 )+({\bf k}-{\bf k}_0 )\cdot
\frac{\partial q_z^H({\bf k})}{\partial {\bf k}} \bigg|_{{\bf
k}_0}.\label{te}
\end{equation}
The complex field in region $3$ can  be obtained simlarlly
\begin{equation}
E_{3}(x, z) = \int_{-\infty}^{+\infty}d k_\perp f( k_\perp )T_2
({\bf k}) \exp[i({\bf k}_0+{\bf k}_\perp) \cdot {\bf r}- i
\omega_0  t].\label{FIII}
\end{equation}
where $T_2({\bf k})$ is the transmission coefficient of the
modulated Gaussian beam at the second interface. E- and
H-polarized waves can totally propagate thought the interface.

To this end, the intensity distribution of the transmission field
in the free space and anisotropic media slab can be derived from
Eqs.~(\ref{FI}), (\ref{FII}) and (\ref{FIII}) under the above
approximation. For simplicity, the isotropic medium is assumed to
be a vacuum in our calculation. Note that the refraction angles of
energy of E- and H-polarized beams in the anisotropic medium slab
are almost exactly the analytical expression in Eq.~(\ref{AS}).
Our numerical results indicate that it is advantageous to employ
the anisotropic media slab as polarization beam splitter.

Finally we want to enquire: how can the beam splitting effect be
studied experimentally? Generally speaking, it is not strange to
mention the question because no scheme can be of much interest if
the means of realizing it are not available. Fortunately several
recent developments make the beam splitter a practical
possibility. An extremely promising material has been previously
explored in certain designs of photonic crystals, which can be
effectively modelled with anisotropic permittivity and
permeability tensors~\cite{Shvets2003,Shvets2004,Urzhumov2005}.
Therefore there is no physical and technical obstacle to construct
the anisotropic medium slab to split polarized beams.

\section{Conclusion}\label{sec4}
In conclusion, we have investigated the wave propagation in the
anisotropic media with single-sheeted hyperboloid dispersion
relation. In such anisotropic media E- and H-polarized waves have
the same dispersion relation while E- (or H-) polarized beam is
positively refracted whereas H- (or E-) polarized beam is
negatively refracted. By suitably using propagation properties in
such kind of anisotropic medium, it is possible to realize very
simple and very efficient beam splitter. We show that the
splitting distance between E- and H-polarized beam is the
functions of anisotropic parameters, the incident angle and the
slab thickness. We are sure that our scheme has not exhausted the
interesting possibilities. In particular, it might be imagined
that a series of beam splitter could be constructed in more
sophisticated processes than those considered here.

\begin{acknowledgements}
This work is supported by projects of the National Natural Science
Foundation of China (No. 10125521 and No. 10535010).
\end{acknowledgements}

\end{document}